   \documentclass[12pt,preprint]{aastex}

\begin{document}

\title{Spectroscopic analysis of interaction between an EIT wave
and a coronal upflow region}

\author{F. Chen, M. D. Ding, P. F. Chen}
\affil{Department of Astronomy, Nanjing University, Nanjing 210093, China}
\affil{Key Laboratory for Modern Astronomy and Astrophysics (Nanjing University), Ministry of Education, Nanjing 210093, China}
\email{dmd@nju.edu.cn}

\author{L. K. Harra}
\affil{UCL-Mullard Space Science Laboratory, Holmbury St. Mary, Dorking, Surrey, RH5 6NT, UK}


\begin{abstract}
We report a spectroscopic analysis of an EIT wave event that occurred in
active region 11081 on 2010 June 12 and was associated with an M2.0
class flare. The wave propagated near circularly. The south-eastern
part of the wave front passed over an upflow region nearby a magnetic
bipole. Using EIS raster observations for this region, we studied the
properties of plasma dynamics in the wave front, as well as the
interaction between the wave and the upflow region. We found a weak
blueshift for the \ion{Fe}{12} $\lambda$195.12 and \ion{Fe}{13}
$\lambda$202.04 lines in the wave front. The local velocity along the
solar surface, which is deduced from the line of sight velocity in the
wave front and the projection effect, is much lower than the typical
propagation speed of the wave. A more interesting finding is that the
upflow and non-thermal velocities in the upflow region are suddenly
diminished after the transit of the wave front. This implies a
significant change of magnetic field orientation when the wave passed.
As the lines in the upflow region are redirected, the velocity along the
line of sight is diminished as a result. We suggest that this scenario
is more in accordance with what was proposed in the field-line stretching
model of EIT waves.
\end{abstract}


\section{Introduction}
Diffuse coronal waves were first observed by \citet{mos97} and \citet{tho98} with the EUV Imaging Telescope (EIT; \citealt{dela95}) aboard the {\it Solar and Heliospheric Observatory} ({\it SOHO}), and were commonly known as EIT waves. They are best seen in the running difference images as propagating bright fronts with a speed of a few hundreds of km s$^{-1}$, followed by an expanding dimming region \citep{tho98}. EIT waves can be observed at several wavelengths, such as 171 \AA, 195 \AA, 284 \AA, and 304 \AA~\citep{wil99,zhu04,lon08}. Many observational properties of EIT waves were presented by \citet{del99}, \citet{kla00}, and \citet{tho09}. Recent reviews on this topic can be found in \citet{wil09}, \citet{war10}, and \citet{gall10}.

It is natural that EIT waves are accompanied with some other solar active events such as coronal mass ejections (CMEs) and flares. Statistical and case studies have shown that EIT waves are closely related with CMEs rather than solar flares \citep{bie02,oka04,chen06}. In particular, the flares associated with many EIT wave events were very weak \citep{cliv05,ver08,ma09,att09}. In spite of the coincidence of EIT waves and CMEs, their spatial relationship is still being debated. \citet{vrs06} reported a wave behind the CME flank, and \citet{pat09} claimed that the EIT wave front is outside the CME frontal loop, whereas \citet{chen09} and \citet{dai10} found that the EIT wave front is cospatial with the white-light frontal loop of CMEs. In addition, \citet{war10} found one event in which a wave front overtakes the CME flank.

Based on the observed properties of EIT waves, there are two main interpretations. One is the wave model, which suggests that EIT waves are fast-mode magnetohydrodynamic (MHD) waves or shocks \citep{wang00,wu01,war01,war04,pom08,vrs08}. This model can explain some of the characteristics of the wave front and was supported by a number of observations \citep{war05,lon08,ver08,pat09,gopa09,kie09}. The other interpretation is related to CME expansion. \citet{del00} suggested that the bright front may result from the interaction between CME-induced expansion of magnetic field lines and surrounding field lines. \citet{chen02,chen05} proposed a field-line stretching model, which predicts that there should exist a fast moving coronal shock ahead of the slow EIT wave, which were recently confirmed by \citet{chen11}. The non-wave model can also explain many of the properties of the wave and was supported by \citet{har03} and \citet{zhu09}. In addition, there are still other models such as slow-mode MHD waves \citep{wang09}, spherical current shell \citep{del08}, successive magnetic reconnections \citep{att07a,att07b}, and soliton \citep{wil07}. A concept of a coupled coronal wave was proposed by \citet{zhu04} and \citet{coh09}.

In the last decade, the main approach to studying EIT waves is by ultra-violet imaging observations. The main limitation comes from the cadence and spatial resolution of the instruments, such as EIT on board the {\it SOHO} and Extreme UltraViolet Imager (EUVI) on board the {\it Solar TErrestrial RElations Observatory} ({\it STEREO}; \citealt{how08}). Atmospheric Imaging Assembly (AIA; \citealt{tit06}) aboard the {\it Solar Dynamics Observatory} ({\it SDO}) has high spatio-temporal resolutions and signal-to-noise ratio, enabling us to study EIT waves in unprecedented detail. \citet{liu10} reported the first {\it SDO}/AIA observations of EIT wave. They found one diffuse pulse and multiple sharp fronts and suggested a hybrid interpretation, combining both wave and non-wave models. In addition to the imaging observations, spectroscopic observations are also very important, because they provide additional information on plasma dynamics during the wave propagation and aid  clarification of the physical nature of EIT waves. \citet{har03} did the first spectroscopic analysis of EIT waves using Coronal Diagnostic Spectrometer (CDS; \citealt{harrison95}) on board the {\it SOHO} spacecraft. They measured Doppler velocities in the wave front and the following dimming. They found an absence of Doppler velocity in the wave front but an ejection of cold material after the wave front passed. Using the EUV Imaging Spectrometer (EIS; \citealt{cul07}) on {\it Hinode}, \citet{asa08} studied a fast mode MHD shock wave visible in soft X-rays. Unfortunately, some of the EIT images suffered from scattered light in the telescope. Therefore, the wave front in EIT data was unclear. More recently, \citet{chen10} confirmed the studies of \citet{har01,har03}, and further found an enhanced line broadening at the outer edge of the dimming region, which could be well explained by the field-line stretching model of \citet{chen02,chen05}.

In this paper,we present a case study of {\it Hinode}/EIS observation of an EIT wave event. We successfully obtained the temporal evolution of the line intensity, line width, and Doppler velocity for two iron lines in a sliced region overlapping a small upflow region. Hence, we are able to reveal the interaction between the EIT wave and the coronal upflow region spectroscopically for the first time. We describe the observations and data analysis in \S \ref{obda}. Our results are shown in \S \ref{result}, followed by some discussions on the results in \S \ref{discu}.

\section{Observations and Data Analysis}\label{obda}
\begin{figure}
\epsscale{1}
\plotone{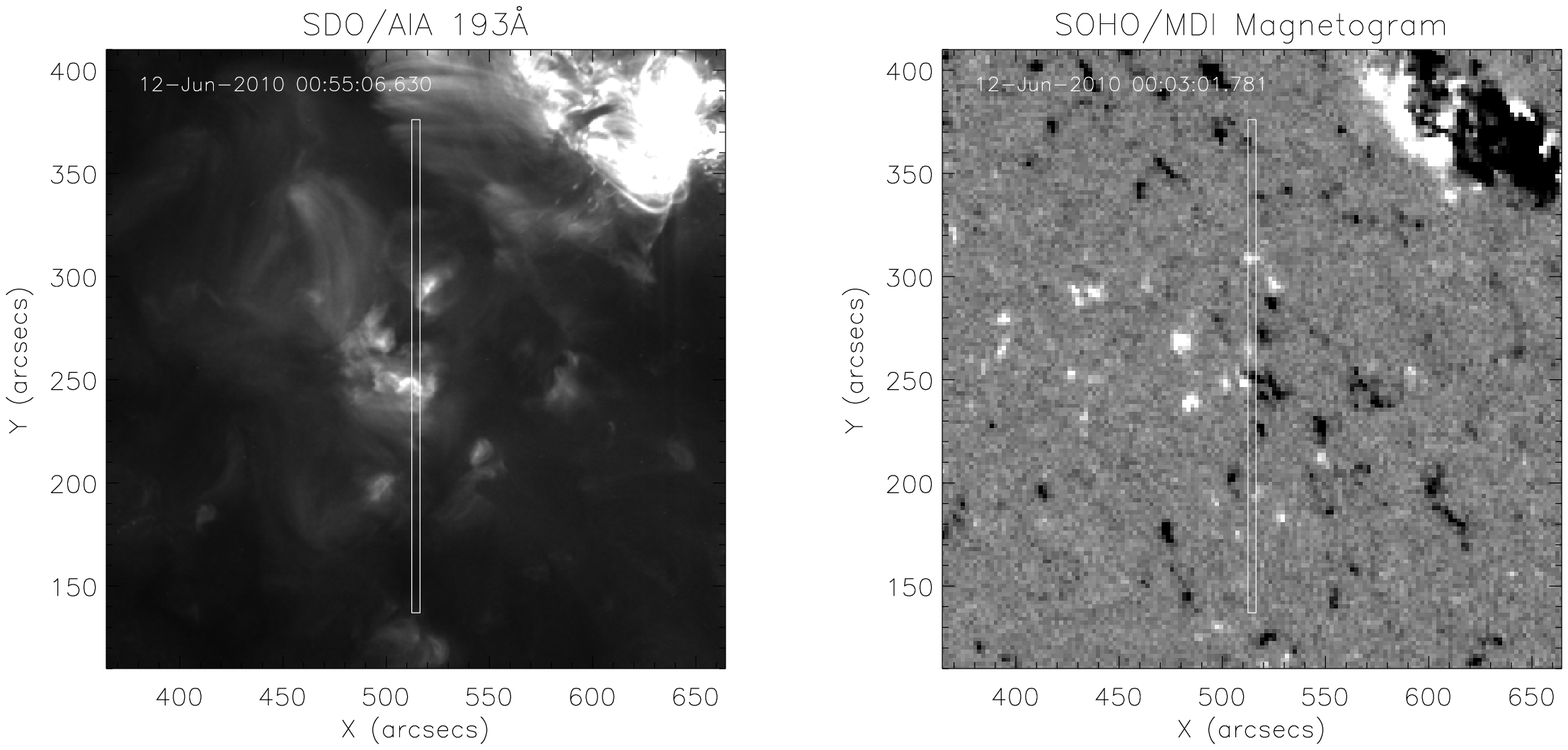}
\caption{SDO/AIA 193 \AA~image and the magetogram measured by the Michelson Doppler Imager (MDI; \citealt{sche95}) aboard {\it SOHO}. The white boxes indicate the EIS FOV.}\label{mag}
\end{figure}

The EIT wave event we studied occurred in the active region 11081 on 2010 June 12 and was associated with an M2.0 class flare. The EIS observations started at $\sim$~00:35 UT, using a 1\arcsec~slit with a step of 1\arcsec~and an exposure time of 60 s. The time gap between successive exposures is 2 s. The field of view (FOV) is 5\arcsec~in the scanning direction and 240\arcsec~in the slit direction. Therefore, a raster cadence of $\sim$~310 s for EIS was achieved. This raster was repeated 12 times. The FOV of EIS lay to the south of the active region where the flare occurred and the wave was generated. The FOV covered the central part of a magnetic bipole as shown in Figure \ref{mag}.

\begin{figure}
\epsscale{1}
\plotone{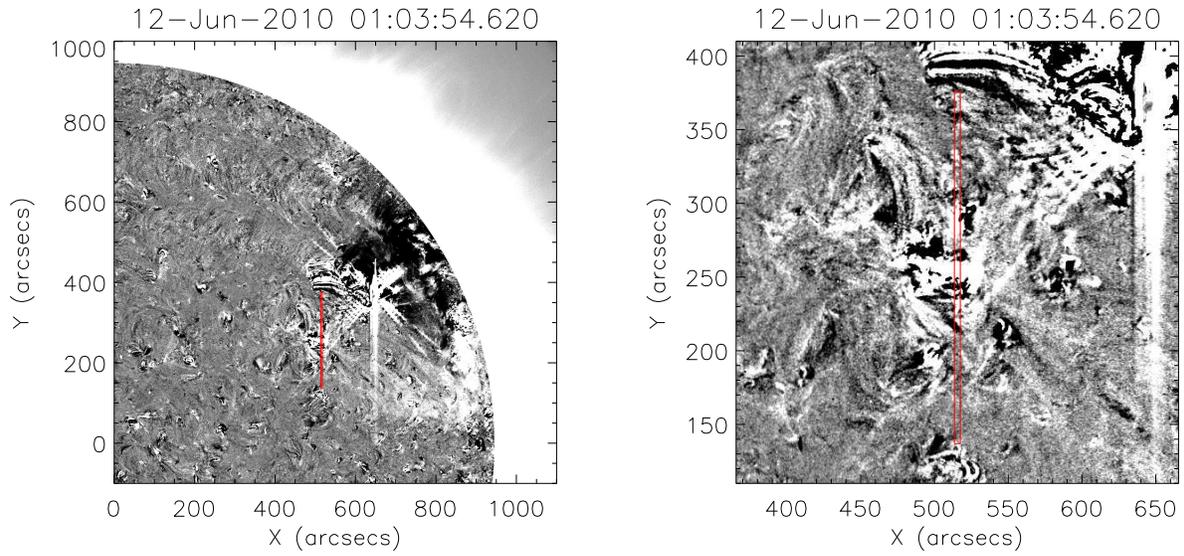}
\caption{Base difference image of SDI/AIA 193 \AA. The image is subtracted by the one at 00:55:06 UT, with the solar rotation corrected. The red boxes indicate the EIS FOV. [This figure is also available as an mpeg animation in the electronic edition of the article.]}\label{snapshot}
\end{figure}

\begin{figure}
\epsscale{1}
\plotone{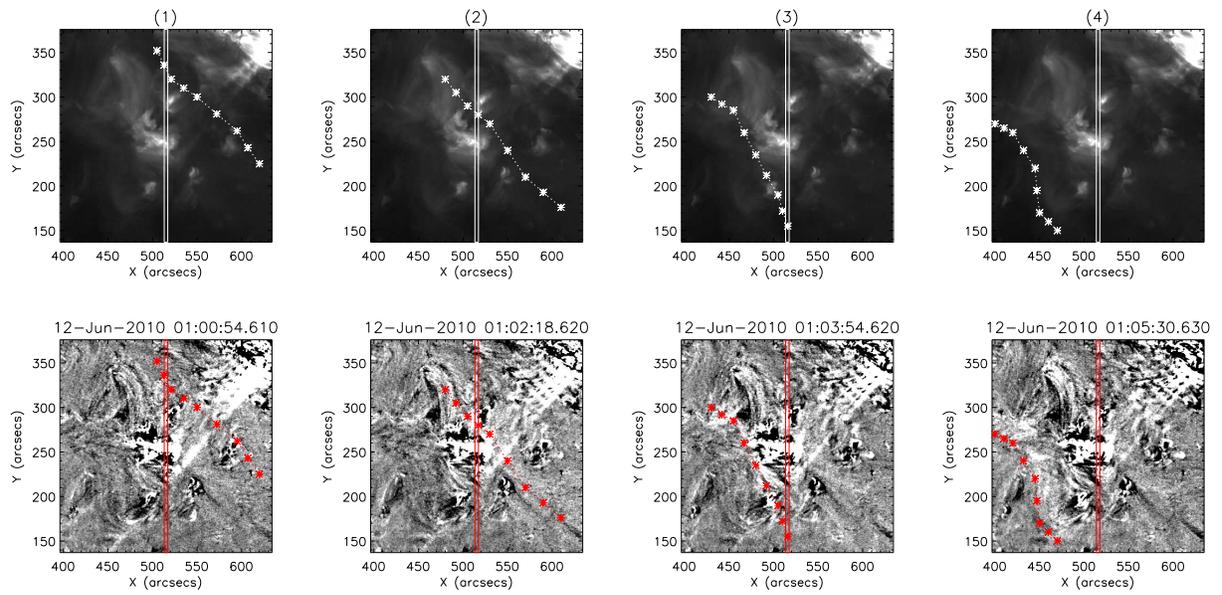}
\caption{Positions of the wave front indicated by the asterisks measured from SDO/AIA base difference images. The white/red boxes refer to the EIS FOV. Note that the wave front may not appear clear enough in some snapshots, in which the position are measured using movie instead.}\label{wave}
\end{figure}

This event was also well observed by {\it SDO}/AIA in several coronal passbands. We obtained the base difference image from the 193 \AA~data between $\sim$~00:55 and $\sim$~01:15 UT and found that the wave propagated nearly circularly, as shown in Figure \ref{snapshot}. Fortunately, the south-eastern part of the wave front, though not the strongest, passed through the EIS FOV. For a more detailed study, it is important to determine the spatial intersection between the wave and EIS observations. Thus, we measured the positions of the leading edge of the wave front with the help of AIA 193 \AA~base difference movie and marked the results on both the AIA 193 \AA~maps, as shown by the asterisks in Figure \ref{wave}. The wave front entered the EIS FOV at $\sim$~01:00 UT and moved out of it at $\sim$~01:04 UT. Although the cadence of $\sim$~310 s is a relatively high one for EIS observations, it is still too low to observe the wave with good temporal resolution. In this event, only one of the 12 rasters, which started at 01:01:02 UT, may contain some signatures of the wave front. There is usually an expanding dimming region behind the wave front; however, the dimming to the south-east of the active region was not obvious. It is possible that the dimming region did not propagate circularly as the wave front in this event. Another possibility is that the dimming was obscured by the presumed overlying dome-like wave front that contributed a part of the EUV emission.

For the analysis of EIS data, we selected the \ion{Fe}{12} $\lambda$195.12 and \ion{Fe}{13} $\lambda$202.04 emission lines, since both of them are strong with no blends from other ions \citep{you07,you09}. In addition, there are no significant line asymmetries, since the EIS FOV is mainly in a quiescent region rather than in an active region \citep{pet10}. Hence we used a one-component Gaussian function to fit the line profiles. The function we used is written as
\begin{equation}\label{func}
I=A+B\lambda+I_{0}\exp\left[-\frac{(\lambda-\lambda_{0})^{2}}{2\sigma^{2}}\right],
\end{equation}
where $\lambda$ (in units of \AA) is the wavelength, $I_{0}$ (in units of ergs~cm$^{-2}$~s$^{-1}$~sr$^{-1}$~\AA$^{-1}$) is the peak value of the profile, $\lambda_{0}$ (in units of \AA) is the line center, $\sigma$ (in units of \AA) is the line width, and $A$ and $B$ are constants for the linear background.

The EIS FOV in the scanning direction is very small (5\arcsec), thus our main interest focuses on the variation along the slit. Hence, we binned up the 5 pixels along the scanning direction for all the rasters and obtained 12 slices. Figure \ref{eis} shows the synthetic maps reconstructed from the 12 slices for the \ion{Fe}{12} and \ion{Fe}{13} lines. The positions of the wave front measured from AIA images are plotted in each panel with the dashed lines indicating the wave propagation. The detailed results of spectroscopic analysis are presented in the following section.

\section{Results}\label{result}

\subsection{The properties of the wave front}\label{result_front}

It is of interest to determine the physical characteristics of EIT waves from the spectroscopic observations. However, the cadence of scanning is usually too low to keep in pace with a fast propagating  phenomenon. Hence, it is challenging to observe the wave front in the line intensity maps reconstructed from the raster scans \citep{chen10}. In this event, we cannot see an obvious intensity increase along the dashed line in Figure \ref{eis}, which indicates the wave propagation. Usually, we may find some distinct features in line widths and Doppler velocities as revealed in previous studies \citep{har03,asa08,chen10}. As shown in Figure \ref{eis}, the variation of line width along the wave propagation in the quiescent region is insignificant and within the fitting error. This is different from the result observed by \citet{chen10}, in which an enhanced line broadening appeared at the outer edge of the ensuing dimming, i.e., behind the wave front.

\begin{figure}
\epsscale{1}
\plotone{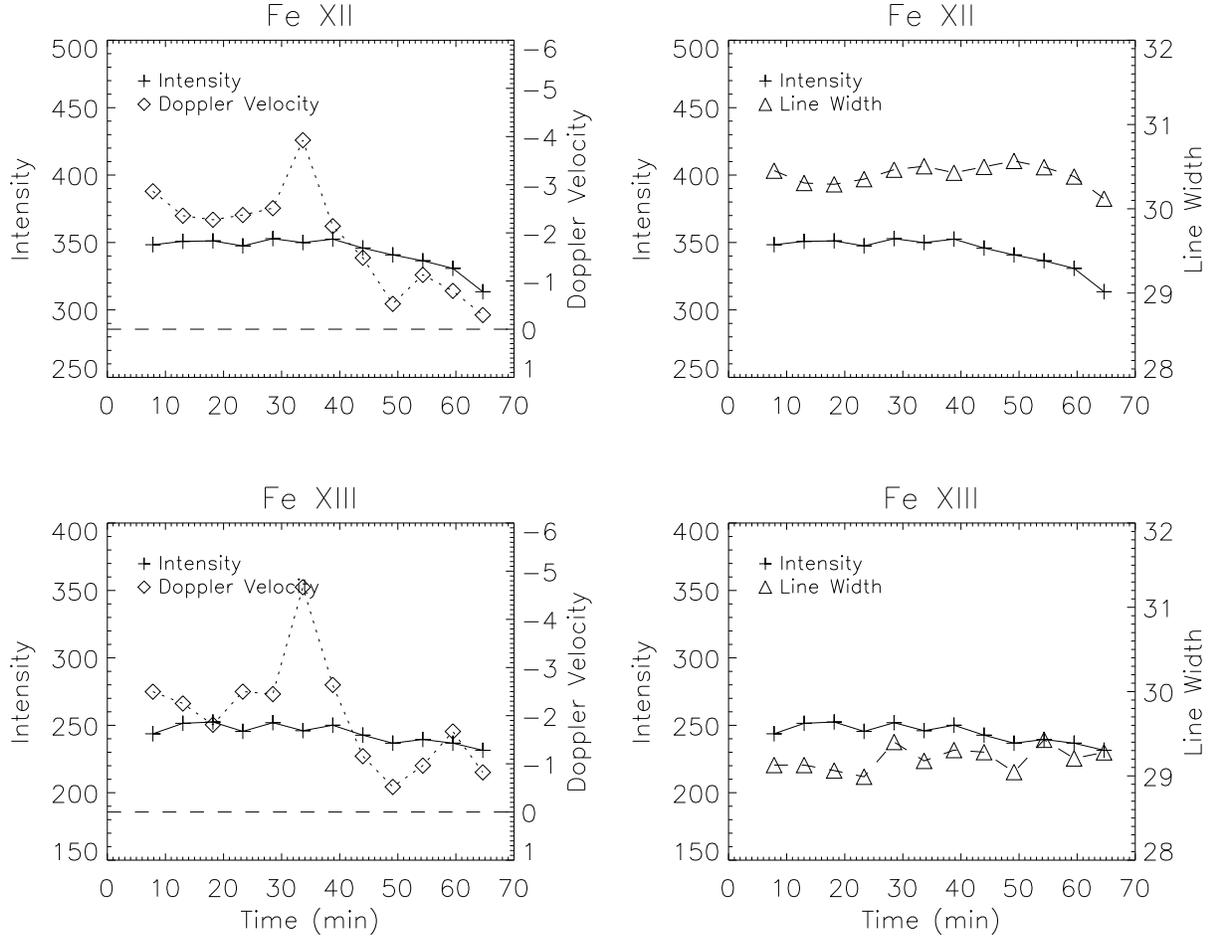}
\caption{Average line intensity (in units of ergs cm$^{-2}$~s$^{-1}$~sr$^{-1}$), line width (in units of m\AA), and Doppler velocity (in units of km s$^{-1}$) between Y=330\arcsec~and Y=364\arcsec~as a function of time. See the legend for details. Negative values are for blueshifts. The dashed lines in the left column indicate 0 km s$^{-1}$. The time is related to 00:30:00 UT.}\label{blueshift}
\end{figure}

Unlike the wave-induced effect on the line intensity and line width, we discern a weak blueshift in lines along the wave propagation in the upper part of the EIS FOV. The blueshift is stronger for the \ion{Fe}{13} line. To illustrate the result, we calculated the average of the Doppler velocities between Y=330\arcsec~and Y=364\arcsec~for each slice. The average Doppler velocity, as well as other line parameters, is plotted as a function of time in Figure \ref{blueshift}. We can confirm that the line intensity and line width in this region did not vary significantly with time. At the time of the wave front transit (i.e., $\sim$~33 min according to Figure \ref{blueshift}),  the blueshift for the \ion{Fe}{12} line suddenly increased to $\sim$~4 km s$^{-1}$, a value that was different from those observed earlier by nearly 2 km s$^{-1}$. For the \ion{Fe}{13} line, the amplitude of the blueshift when the wave passed was a little larger than that for the \ion{Fe}{12} line. The average fitting error in this region was $\sim$~1.41 km s$^{-1}$. The blueshift, albeit weak, was beyond the errors.

\subsection{Interaction between the EIT wave and the upflow region}\label{result_upflow}
The EIS FOV covered the central part of a magnetic bipole, which appeared as several minor loops and EUV bright points in AIA images as shown in Figure \ref{mag}. The photospheric magnetic field strength was generally less than 100 Gauss. In Figure \ref{wave}, we found that the wave front was only slightly distorted when it passed over these magnetic structures. It is known that the EIT wave front usually stops at active region boundaries and coronal holes \citep{tho98,ver06,ver08}. The magnetic bipole here may be too small to stop the wave propagation, in either wave scenario or non-wave scenario.

There was an upflow region to the north of the magnetic bipole core (Y $\sim$~245\arcsec) that appeared the brightest in both the AIA image and the EIS line intensity map. In the upflow region, the line intensities of both the iron lines were much lower. The velocity amplitude of the upflow was 15 km s$^{-1}$ for the \ion{Fe}{12} line and 20 km s$^{-1}$ for the \ion{Fe}{13} line. The line width in this upflow region was larger than that in the ambient regions. Moreover, the non-thermal velocity ($v_{non}$) can be calculated by
\begin{equation}\label{non}
{\rm FWHM}_{obs}^{2}={\rm FWHM}_{ins}^{2}+4{\rm ln}2~\frac{\lambda^2}{c^2}~\left(\frac{2kT}{M}+v_{non}^2\right),
\end{equation}
where the value of FWHM$_{ins}$ is 0.056 \AA, $\lambda$ is the wavelength (in units of \AA), $c$ is the speed of light, $k$ is the Boltzmann constant, $T$ is the electron temperature, and $M$ is the ion mass. The value of FWHM$_{obs}$ (in  units of \AA)  can be obtained from the line profile. In Figure \ref{eis}, the upflow region was clear in the Doppler velocity and line width synthetic maps for both the iron lines. The upflow velocity and line width stayed nearly unchanged until the wave front passed. However, after the wave front passed the upflow region, we found that the upflow velocity and line width to the right of the dashed line suddenly and significantly decreased for both the lines. We refer to this phenomenon as a diminishing of upflow and non-thermal velocities. For a quantitative study, the Doppler velocity in this region is plotted against the non-thermal velocity, as shown in Figure \ref{velocity}. It is clear that after the wave front passed, the high velocity tail was truncated so that the upflow velocity was $\la$~5 km s$^{-1}$ and the non-thermal velocity was $\la$~70 km s$^{-1}$ for the \ion{Fe}{12} line; while the upflow velocity was $\la$~10 km s$^{-1}$ and the non-thermal velocity was $\la$~60 km s$^{-1}$ for the \ion{Fe}{13} line.

\begin{figure}
\epsscale{1}
\plotone{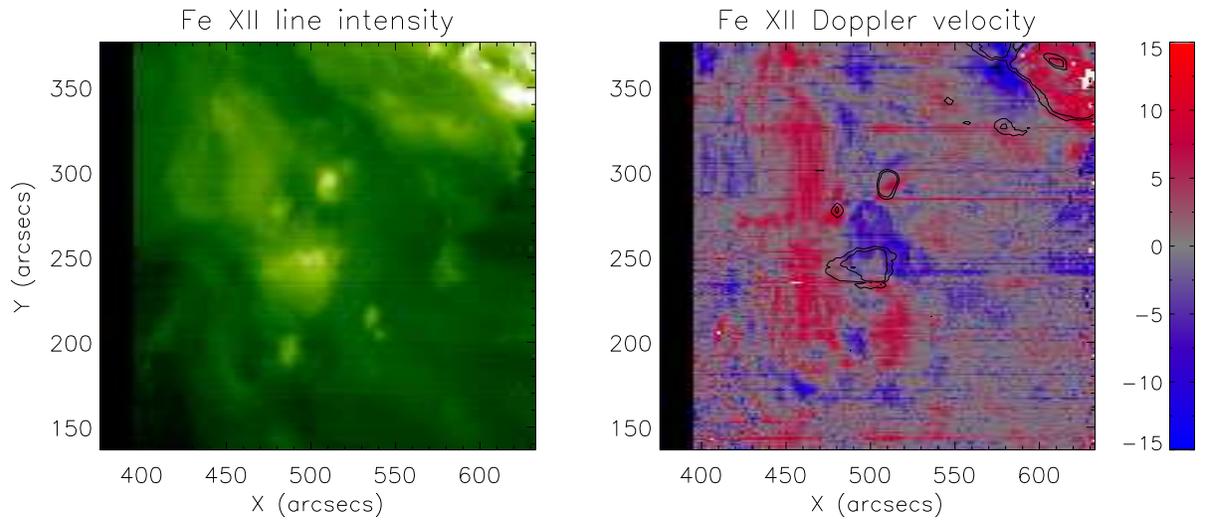}
\caption{Line intensity and Doppler velocity maps obtained from EIS raster from 22:57:40 UT 2010 June 11 to 00:05:13 UT 2010 June 12. The Doppler velocity map is overlaid by the contours of the line intensity, which show the position of the  magnetic bipole core. Shown in the right is the color bar for the Doppler velocity.}\label{preraster}
\end{figure}

In addition, we analyzed the data obtained with a long raster (a large FOV) on the day before the event. The line intensity and Doppler velocity maps are shown in Figure \ref{preraster}. We can confirm that the upflow region existed at this position and exhibited the similar velocity amplitude for many hours. The magnetic bipole core is illustrated by the black contours on Figure \ref{preraster} (the right panel). The abrupt diminishing of the upward Doppler velocity and the non-thermal velocity in an ambient region of magnetic structure may imply an interesting scenario, in which the wave front interacts with the magnetic field during its propagation.

\section{Discussion}\label{discu}

We presented an EIT wave event, which was captured by {\it Hinode}/EIS raster. We studied the spectroscopic properties of the coronal plasma during the wave transit. The most interesting results are summarized below.

\begin{enumerate}
\item{The wave front propagated near circularly, with its south-eastern part observed by EIS. The line intensity and line width showed no change when the wave front passed. However, an enhanced blueshift of the line center, albeit weak, was observed at this time.}
\item{The wave front passed over an upflow region nearby a magnetic bipole. The shape of the wave front was only slightly distorted. For both of the iron lines studied, a sudden and clear diminishing of the upflow and non-thermal velocities in the upflow region was observed when the wave passed.}
\end{enumerate}

In general, during the EIT wave front passage, one may expect to observe a line intensity enhancement, a common feature observed in coronal EUV images.  In this event, however, the variation of line intensity during the wave propagation was within the fitting error. For the raster observations of the wave, the line profile at a certain point is a composite of the contribution from the background corona during the whole exposure time (50 s for this event) and the contribution from the wave front. The wave quickly passes over this point and in fact it contributes little to the line profile. Therefore, it is hard to observe similar signatures from spectroscopic and image observations. It was previously observed that EIT waves and the accompanied dimmings exhibited a very high speed, so that its contribution to the line profile can result in a strong Doppler component \citep{har03,asa08}. In this case, some significant features of the wave or wave-perturbed plasma can be observed from spectroscopic data.

Fortunately, we found a sudden upward motion of the coronal plasma when the wave front passed the EIS FOV. The upward velocity ($v_{up}$) was $\sim$~4 km s$^{-1}$. If we assume that the velocity along the line of sight was purely from the projection of a tangential velocity (i.e., velocity along the solar surface), then we can estimate the local tangential velocity to be $v_{up}$/$\cos\theta$, where $\theta$, the projection angle, is $\sim$~30$\degr$. Thus, the local tangential velocity was $\sim$~10 km s$^{-1}$. For this event, the average propagation speed of the wave between 1:00:54 and 1:05:30 UT is $\sim$~358 km s$^{-1}$. This result implies that the speed of local plasma motion was far less than the propagation speed of the wave. This feature is compatible with both the wave models and non-wave models.

The upflows at the edge of the active regions have been reported by \citet{sakao07}, \citet{hara08}, \citet{har08}, and \citet{dosc08}. They were found to be cospatial with open magnetic field lines \citep{sakao07,har08}. The upflow region observed in our event, as described in Section \ref{result_upflow}, had some similar behaviors. However, its spatial extension was much smaller. Base on the fact that the upflow we observed existed in a low intensity region, it is probable that this region was a small coronal hole. Regardless whatever the actual source of the upflow was, the magnetic field lines in this region were very likely to be open. However, several difficulties make it hard to get a valid result on the magnetic field structure for this event. First, both the active region where the EIT wave originated and the magnetic bipole region where upflows existed were far from the disk center. A correction for the curved surface and the projection of magnetic fields is required. Second, although extrapolation to active region magnetic fields has been intensively studied and relatively easy to apply to various cases, extrapolation to small magnetic structures is a difficult task. Unfortunately, the region with the magnetic bipole and the upflows, which draws our main attention, is such an example. Our experiment showed that it is possible to get some preliminary results, though probably inaccurate, for the active region. However, no valid result is returned for the magnetic bipole or the intermediate area between it and the active region. Although there is no available information on the magnetic fields, we conjecture that the upflow region surrounding the magnetic bipole core may also be related with open fields or large-scale close fields, based on the reasons mentioned above.

The ratio of magnetic pressure to gas pressure is high in the corona; thus, the direction of coronal plasma motion is dominated by magnetic field lines. The sudden variation of the line of sight velocity described in Section \ref{result_upflow} implies the change of direction of magnetic field lines. \citet{mcin07} reported a disappearance and reappearance of ``moss" around an active region during the evolution of a CME event. They explained the change of the ``moss" as a proxy of the changing coronal magnetic field topology behind the CME front. Although their event is different from ours in some aspects, the common key feature is that a change of magnetic field orientation associated with large scale coronal disturbances (CMEs or EIT waves) can influence the coronal plasma dynamics. In the field-line stretching model proposed by \citet{chen02,chen05}, EIT waves are apparently-propagating wave fronts formed by successive stretching/expansion of field lines, which is initiated by the erupting flux rope. It is expected that when the wave passed over the upflow region, the local open (or large-scale) field lines would be pushed aside and be redirected toward another direction by the stretched field lines related with the wave front. Therefore, the upflows changed their direction to that deviated more from the line of sight. The line of sight component velocity were diminished apparently as a result.

\begin{figure}
\epsscale{1}
\plotone{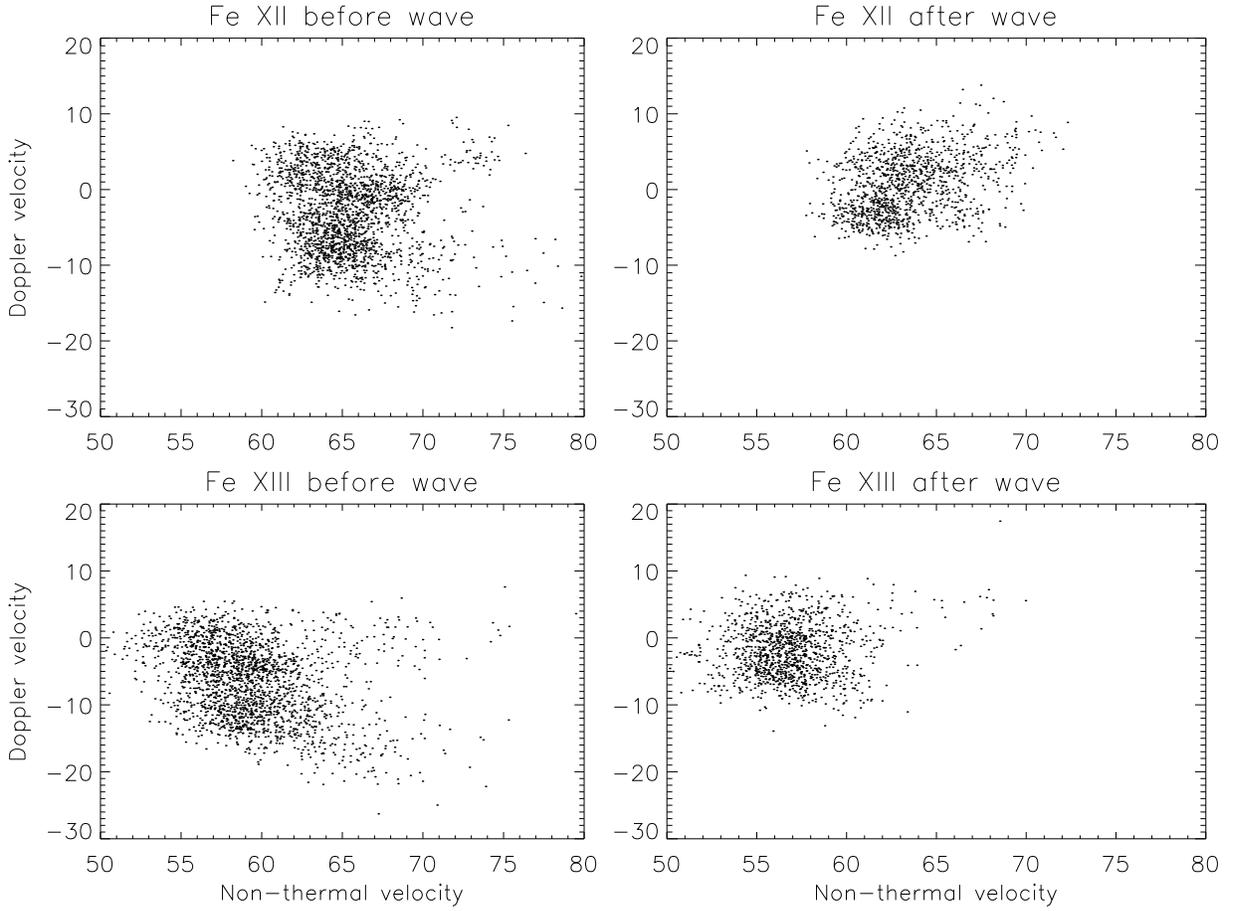}
\caption{Scatter plot of the Doppler velocities (in units of km s$^{-1}$) against the non-thermal velocities (in units of km s$^{-1}$) obtained from line widths for pixel points in the upflow region.}\label{velocity}
\end{figure}

\citet{hara08} and \citet{dosc08} reported a positive correlation between the non-thermal velocity and upflow velocity from the line profile fitting, implying the multiplicity of actual upflow velocities. As shown in Figure \ref{velocity}, before the wave front passed, the upflow velocity and non-thermal velocity in the region we observed also shows a positive correlation, though somewhat weaker than what was reported by \citet{hara08} and \citet{dosc08}. However, such a correlation does not exist after the wave front passed, when the upflow and non-thermal velocities were reduced to smaller magnitudes as mentioned in Section \ref{result_upflow}. Considering the difference in the correlation result, we think that the correlation observed in this upflow region before the wave front passed is of physical significance, especially for the \ion{Fe}{13} line. Since one of the main causes of the non-thermal velocity is the multiplicity of line of sight velocities, the sudden diminishing of the non-thermal velocity could therefore be explained. Note that other mechanisms were also suggested to explain the variation of non-thermal velocity in coronal dimmings \citep{mcin09}. Here, we favor the scenario of multiple velocities existing in the spatially unresolved area to interpret the variation of the non-thermal velocity.

\begin{figure}
\epsscale{1}
\plottwo{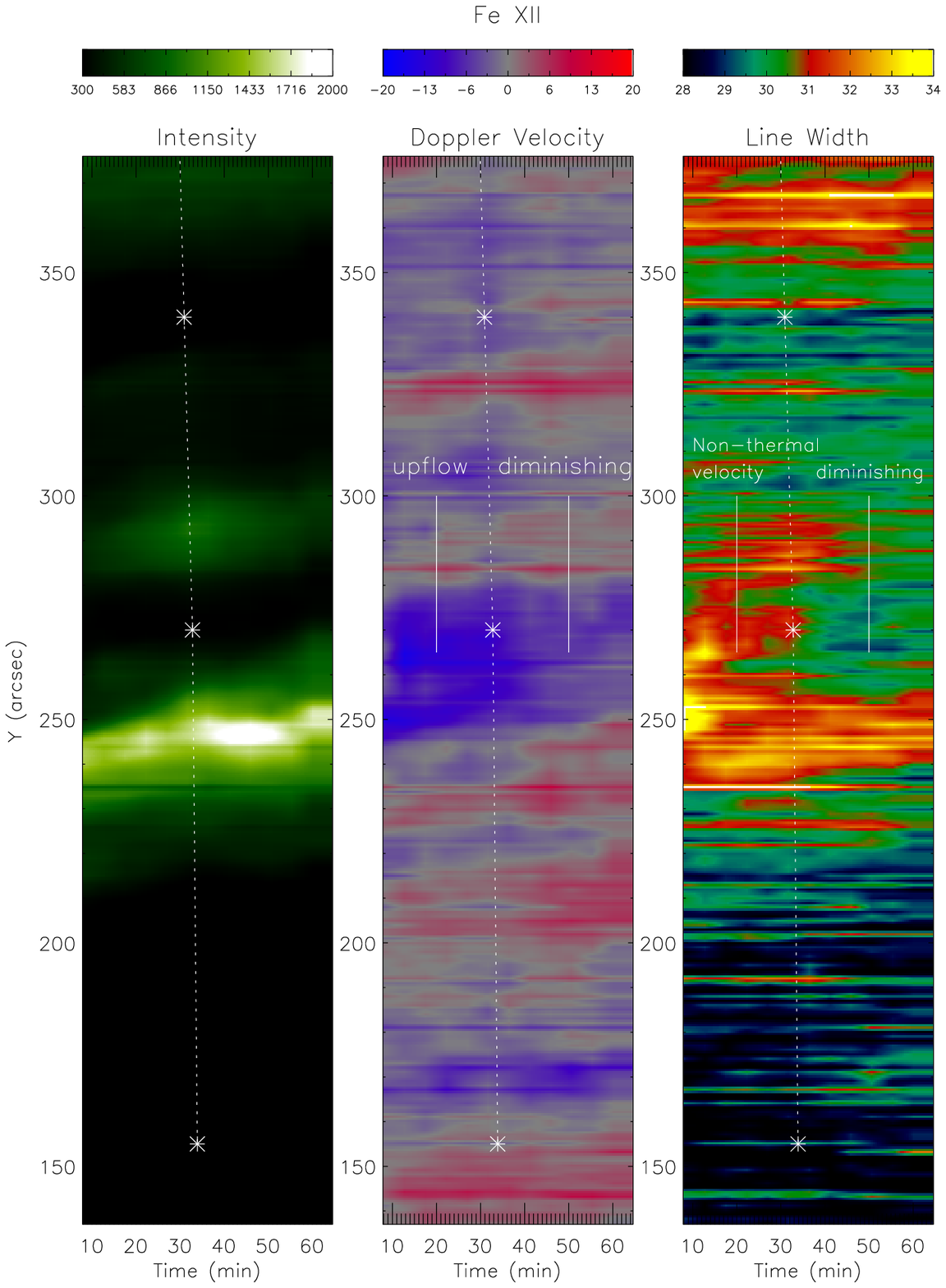}{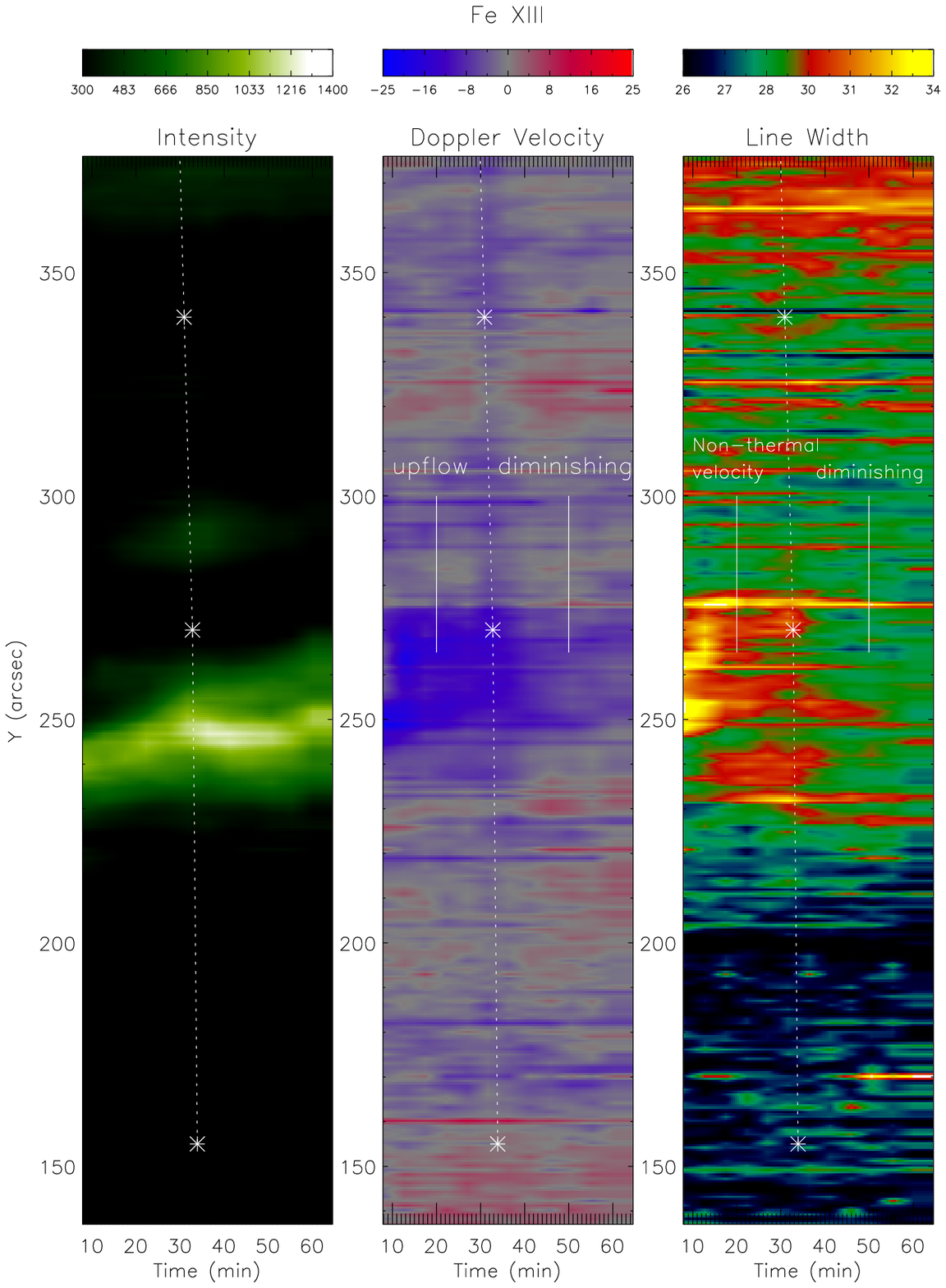}
\caption{Line intensity (left), Doppler velocity (middle), and width (right) for the \ion{Fe}{12} and \ion{Fe}{13} lines as a function of time. The time is related to 00:30:00 UT. The asterisks indicate the positions of the wave front measured from Figure \ref{wave}. Note that in the fourth colume of Figure \ref{wave}, the wave front is already out of the EIS FOV. The dashed lines connecting the asterisks indicate the wave propagation track.}\label{eis}
\end{figure}

Note that an MHD wave impacting open (or large-scale) field lines can
also push them aside and redirect them to some extent. However,
the field lines would oscillate periodically.
If so, we expect to observe some oscillating
patterns in the spectroscopic data. Unfortunately, no oscillation
is found in the results shown above. It may be true that the
sensitivity and the cadence of the instruments are not high enough
to resolve the oscillation well. If this is the case, the
diminishing of the Doppler velocity and non-thermal velocity
would be gradual. However, Figure \ref{eis} shows a sudden diminishing
of the velocities for both the iron lines. Therefore, while the wave
models cannot be excluded, the non-wave model is more favored here in
explaining the observational data.

In summary, a possible scenario for the EIT wave event analyzed in this
paper could be as follows: the EIT wave encountered an open (or large-scale)
magnetic field line, along which multi-component upflows
existed. The independent magnetic system of the upflow region was too
narrow to stop the wave from propagating. However, the open (or
large-scale) field lines were pushed aside or distorted by the
wave, resulting in an abrupt diminishing of upward line of sight
velocity and non-thermal velocity as observed in spectral lines. We
suggest that this scenario could be more in accordance with
what proposed by \citet{chen02,chen05} for EIT waves, in which
the formation of EIT wave fronts and their behaviors are essentially
correlated with the stretching of magnetic field lines during CMEs.

\acknowledgments
We thank the anonymous referee for constructive comments that helped to improve this manuscript. This work was supported by NSFC (under grants 10828306, 10933003, and 11025314) and by NKBRSF under grant 2011CB811402. {\it Hinode} is a Japanese mission developed and launched by ISAS/JAXA, collaborating with NAOJ as a domestic partner, NASA and STFC (UK) as international partners. Scientific operation of the {\it Hinode} mission is conducted by the {\it Hinode} science team organized at ISAS/JAXA. This team mainly consists of scientists from institutes in the partner countries. Support for the post-launch operation is provided by JAXA and NAOJ (Japan), STFC (U.K.), NASA, ESA, and NSC (Norway). We thank the SDO/AIA consortium for providing open access to their calibrated data.

\clearpage

\clearpage

\clearpage

\clearpage

\clearpage

\clearpage

\clearpage

\end{document}